\documentclass[oneside]{article}
\usepackage{epsfig}

 \usepackage{graphicx}
\usepackage[pdftex]{color}
\usepackage{amsmath}
\usepackage{amssymb}

\newtheorem{theorem}{Theorem}[section]
\newtheorem{corollary}[theorem]{Corollary}
\newtheorem{proposition}{Proposition}
\newtheorem{lemma}[theorem]{Lemma}
\newtheorem{definition}[theorem]{Definition}

\def \beq{\begin{equation}}
\def \eeq{\end{equation}}
\def \beqa{\begin{eqnarray}}
\def \eeqa{\end{eqnarray}}
\def \beqan{\begin{eqnarray*}}
\def \eeqan{\end{eqnarray*}}
\def \bea{\begin{eqnarray}}
\def \eea{\end{eqnarray}}

\newcommand{\supp}{{\it Supp}}
\def \um{\frac{1}{2}}

\newcommand{\tr}{\mathop{\rm tr}}

\newcommand{\Hi}{{\cal H}}
\newcommand{\R}{\mathbb{R}}

\newcommand{\N}{\mathbb{N}}
\newcommand{\Z}{\mathbb{Z}}

\newcommand{\Prob}{\mathbb{P}}

\newcommand{\D}{\mathbb{D}}
\newcommand{\E}{\mathbb{E}}
\newcommand{\Ec}{\mathcal{E}}
\newcommand{\Fc}{\mathcal{F}}
\newcommand{\Rc}{\mathcal{R}}

\newcommand{\qed}{\hfill $\Box$ \vskip 2ex}

\def \trace{\textrm{trace}}

\def \supp{\textrm{supp}}

\def \um{\frac{1}{2}}

\def \diag{\textrm{diag}}

\def \tr{\textrm{trace}}
\def \trace{\textrm{trace}}

\def \proofn {\noindent {\em Proof. }}
\def \qed{\hfill $\Box$ \vskip 2ex}

\begin{document}
\setlength{\textheight}{8.0truein}    


\thispagestyle{empty}
\setcounter{page}{1}


\vspace*{0.88truein}



\centerline{\bf
\uppercase{On Time-reversal and space-time harmonic processes}} 
\vspace*{0.035truein}
\centerline{\bf \uppercase{for Markovian quantum  channels}}
\vspace*{0.37truein}

\centerline{\footnotesize
FRANCESCO TICOZZI\footnote{ticozzi@dei.unipd.it}}
\vspace*{0.015truein}
\centerline{\footnotesize\it Dipartimento di Ingegneria dell'Informazione, Universit\`a di Padova, via Gradenigo 6/B}
\baselineskip=10pt
\centerline{\footnotesize\it 35131 Padova, Italy}
\vspace*{10pt}

\centerline{\footnotesize 
MICHELE PAVON\footnote{pavon@math.unipd.it}}\vspace*{0.015truein}
\centerline{\footnotesize\it Dipartimento di Matematica Pura ed Applicata, Universit\`a di Padova, via Trieste 63}
\baselineskip=10pt
\centerline{\footnotesize\it 35131 Padova, Italy}
\vspace*{0.225truein}

\vspace*{0.21truein}

\abstract{
The time reversal of a completely-positive, nonequilibrium  discrete-time quantum Markov evolution is derived via a suitable adjointness relation. Space-time harmonic processes are introduced for the forward and reverse-time transition mechanisms, and their role for relative entropy dynamics is discussed.
}{}{}

\vspace*{10pt}

\vspace*{3pt}

\section{Introduction}


Quantum operations are one of the main mathematical tools to represent generalized measurements, noisy communication channels, the action of the environment on quantum devices and any other uncertain, quantum-state transformation. The study of quantum operations stems from the work of Kraus \cite{kraus} and it has been extensively developed in the operator-algebraic approach to quantum mechanics, see e.
g. \cite{bratteli}. If the time evolution of a quantum system is given by a sequence of quantum operations, then the associated quantum process satisfies the Markov property, namely the future of the process depends on the past only trough its present state \cite{kummerer}. We are interested in studying the form and the properties of the time reversal of this class of quantum Markov evolution.

Time-reversal of quantum operations have been explicitly related to Quantum Error Correction (QEC) in \cite{barnum-reversing}, where the form of a time-reversal for quantum operations has been first proposed in a particular case (full-rank state), and used to discuss the performance of error-correcting codes. In \cite{viola-IPS}, this partial time-reversal is used to characterize correctable codes. Time-reversal of equilibrium quantum Markov processes has been recently discussed in \cite{crooks} with applications to some thermodynamical models. 

In this paper, we introduce a mathematical framework for discrete-time stochastic processes originating from Nelson's kinematics of diffusion processes \cite{nelson-adjoint,N1}. {\em Time-reversal} for Markovian evolutions entails the Lagrange adjoint with respect to the (semi-definite) inner product induced by the flow of probability distributions. We show that this also holds for finite-dimensional, discrete time quantum Markov evolutions. Hence, the time-reversal of a discrete-time quantum  Markov process appears as a peculiar feature of a kinematical nature that is common to all Markovian equilibrium and non-equilibrium evolutions. 

The main contributions of this paper are the following: (i) We derive the form of the backward quantum operation from a general (space-time) adjointness relation, common to all Markovian evolutions; (ii) We establish this result in the general case, i.e. also for states that are neither full-rank nor stationary; (iii) We explicitly show that the derived backward quantum operation not only reverses the state evolution correctly, but among the possibly infinitely many maps that do so, it also ``preserves'' the information about the forward dynamics. That is, the time reversal of the time-reversal operation returns the original quantum channel on the state support; (iv) We introduce the concept of space-time harmonic process and show their central role in entropic evolutions as detailed below.

Time reversal also plays a crucial role in the solution of certain {\em maximum entropy problems} on path space \cite{pavon-bridges} and in deriving an operator form of the {\em H-theorem} (Theorem \ref{qsubm}) for quantum Markov channels. In the above, a key role is played by a suitable class of {\em quantum space-time harmonic 
processes}. We show that, following the analogy with the classical case, they lead to the {\em Belavkin-Stasevski relative entropy} \cite{B-S, F-K, Hiai-P} and their properties imply its monotonicity of the under completely positive, trace-preserving maps. While the latter result was already known \cite{Hiai-P}, it emerges here as a corollary of the properties of space-time harmonic operator processes in a technically much simpler framework.

The paper is structured as follows. In Section \ref{Nelson} we develop a few basic elements of Nelson's kinematics \cite{N1} for discrete-time processes. This is used in Section \ref{markovch} to derive the reverse-time transition mechanism of a Markov chain via a certain adjointness relation. Section \ref{martingale} is dedicated to classical space-time harmonic functions as introduced by Doob \cite{doob-markov} and to the associated {\em martingales}. 
In Section \ref{Qrev}, we proceed to build up the time-reversal of a quantum Markov process by following the same path, i.e. by introducing a suitable space-time semi-definite inner product for quantum observables. Alternative approaches to derive the time-reversal are discussed in Section \ref{other}. Section \ref{Qsth} is devoted to  space-time harmonic processes in the quantum domain and to their key properties.
The remainder of the paper illustrates the role of space-time harmonic processes for information dynamics. After recalling the relevant results for Markov chains in Section \ref{classicalresults}, in Section  \ref{QHtheorem}, {\em Jensen's operator inequality}  allows us to derive an H-theorem in operator form, in striking analogy with the classical case. In Appendix \ref{pathsection}, we sketch some connections to maximum entropy problems on path space and related manifestations of the second law of thermodynamics.

\section{Elements of Nelson's kinematics for discrete-time stochastic processes}\label{Nelson}
Let $I=[t_0,t_1]$ be a discrete-time interval with $-\infty<t_0<t_1<\infty$. Let $(\Omega,{\cal F},\Prob)$ be a probability space and let $\{{\cal F}^-_t\}, t\in I$, be a nondecreasing family of $\sigma$-algebras  of events (filtration) representing a flow of information. Let $X:I\rightarrow L^2(\Omega,{\cal F},\Prob)$ be a second order stochastic process such that $X(t)$ is  $\{{\cal F}^-_t\}$-measurable, for all $t\in I$. Then the {\em conditional forward difference} of $X$ is defined by
$$\Delta^+X(t)=\E(X(t+1)-X(t) | {\cal F}_t^- ).
$$
Consider now a nonincreasing family of $\sigma$-algebras of events $\{{\cal F}^+_t\}, t\in I$, and suppose that $X(t)$ is $\{{\cal F}^+_t\}$-measurable, $\forall t\in I$. Then the {\em conditional backward diference} of $X$ is defined by
$$\Delta^- X (t)=\E(X(t-1)-X(t) | {\cal F}_t^+ ).
$$
Observe that both $\Delta^+X(t), \Delta^- X (t)\in L^2(\Omega,{\cal F},\Prob)$, $\forall t$. A process satisfying $\Delta^+X(t)=0,\forall t\in I$ is called a {\em $\{{\cal F}^-_t\}$-martingale} if $\Delta^+X(t)=0,\forall t\in I$, namely if
\begin{equation}\label{formart}\E(X(t+1) | {\cal F}_t^- )= X(t),\quad a.s.
\end{equation}
It is called a {\em reverse-time, $\{{\cal F}^+_t\}$-martingale} if $\Delta^- X(t)=0,\forall t\in I$, namely if
\begin{equation}\E(X(t-1) | {\cal F}_t^+ )= X(t),\quad a.s.
\end{equation}
If $\Delta^+X(t)\ge0$ or $\Delta^- X(t)\ge 0, \,\forall t\in I$ then $X(t)$ is called a {\em  $\{{\cal F}^+_t\}$ submartingale} and a {\em  $\{{\cal F}^-_t\}$ reverse-time submartingale}, respectively.
We can say that a martingale is {\em conditionally constant} and a submartingale is {\em conditionally increasing}. 
An elementary example of a martingale is provided by the capital of a player at time $t$ in a fair coin tossing game (i.i.d. Bernoulli trials).  The capital is instead modeled by a submartingale if the outcome he is betting on has chance $\ge\frac{1}{2}$ to occur. Notice that the notion of (sub)-martingale is usually defined on the larger class of processes $X$ such that $\E|X(t)|<\infty,\forall t\in I$. Finally, notice that, by iterated conditioning, if $X(t), t\in I$ is a   $\{{\cal F}^-_t\}$-martingale and $Y(t), t\in I$ is a  $\{{\cal F}^-_t\}$-submartingale, then
\beq\label{constexpmart}
\E X(s)=\E X(t), \quad \forall \;s,t\in I,\qquad \E Y(s)\le \E X(t), \quad \forall \;s<t\in I.
\eeq
Similarly, for reverse time (sub)martingales.
A general reference on discrete-time martingales is \cite{neveu}.

Consider now the family ${\cal H}(t_0,t_1)$ of second order stochastic processes $X$ such that $X(t)$ is simultaneously $\{{\cal F}^-_t\}$-measurable and $\{{\cal F}^+_t\}$-measurable, $\forall t\in I$. We then have the discrete-time analogue of Nelson's integration by parts formula \cite[p.80]{N1}.
\begin{theorem} \label{intbyparts}Let $X,Y\in {\cal H}(t_0,t_1)$. Then
\beq
\E\left(X(t_1)Y(t_1)-X(t_0)Y(t_0)\right)=\sum_{t_0}^{t_1-1}\E\left(\Delta^+X(t)Y(t)-X(t+1)\Delta^- Y(t+1)\right).
\eeq
\end{theorem}
\proofn
\beqan &&\E(X(t_1)Y(t_1)-X(t_0)Y(t_0))=
\sum_{t=t_0}^{t_1-1}\E\left[X(t+1)Y(t+1)-X(t)Y(t))\right]=\\
&&\sum_{t=t_0}^{t_1-1}\E\left[(X(t+1)(Y(t+1)-Y(t))
+(X(t+1)-X(t))Y(t)\right].
\eeqan
By the conditional expectation properties, we now have
\beqan \E((X(t+1)-X(t))Y(t))
&=&\E(\E((X(t+1)-X(t))Y(t)| {\cal F}_t^- ))=
\E(\Delta^+X(t)Y(t));\\
\E(X(t+1)(Y(t+1)-Y(t)))
&= &\E(\E(X(t+1)(Y(t+1)-Y(t))| {\cal F}_{t+1}^+ ))
\\&=& -\E(X(t+1)\Delta^- Y(t+1)),
\eeqan
and the conclusion follows.
\qed

\section{Kinematics of Markov chains and space-time harmonic processes}\label{markovch}
Consider a  Markov chain $\{ X(t), t\in \mathbb{Z} \}$ taking values in the finite set ${\cal X}=\{x_1,x_2,\ldots,x_n\}$ which we identify from here on with the set of the indexes $\{1,2,\ldots,n\}$. We denote by $\pi_t$ the probability distribution of $X(t)$ over $\cal X$. In the following, $\pi_t$ is always intended as a column vector, with $i$-th component $\pi_t(i)=\Prob(X(t)=i)$. Let $P(t)$ denote the {\em transition matrix}  with elements $p_{ij}(t)=\Prob(X(t+1)=j|X(t)=i), i,j=1,\ldots,n$. The matrix $P(t)$ is {\em stochastic}, namely
$$p_{ij}(t)\ge 0, \forall i, \forall j, \quad \sum_jp_{ij}(t)=1, \forall i.
$$
Let us agree that troughout the paper $\dag$ indicates  adjoint  with respect to the natural inner product. Hence, in the case of matrices, it denotes transposition and, in the complex case below, transposition plus conjugation. The  evolution is then given by the {\em forward} equation
\beq\label{piev}\pi_{t+1}=P^\dag(t)\pi_t.\eeq
When $P$ does not depend on time, the chain is called {\em time-homogeneous}. A distribution $\bar{\pi}$ is called {\em stationary} for the time-homogeneous Markov chain $X$ with transition matrix $P$ if it satisfies
\begin{equation}\label{stationarity}\bar{\pi}=P^\dag \bar{\pi}.
\end{equation} 
For $x$ and $y$ $n$-dimensional column vectors, we define the semi-definite form:
\beq \label{scalar}\langle x,y\rangle_{\pi_t}=x^\dag D_{\pi_t} y,\eeq 
which is an inner product if $D_\pi=\diag\left(\pi_t(1),\pi_t(2),\ldots,\pi_t(n)\right)$ is positive definite. It represents the expectation of the random variable $Z$ defined on $({\cal X},\pi_t)$ by $Z(i)=x_iy_i$. In what follows, whenever a matrix $M$ is not invertible, $M^{-1}$ is to be understood as the generalized (Moore-Penrose) inverse $M^{\#}$, cf. \cite{horn-johnson}.

\subsection{Space-time inner product and time-reversal}
Let  ${\cal F}^-_t, t\in\mathbb{Z}$ be the $\sigma$-algebra generated by $\{X(s), s\le t\}$ and ${\cal F}^+_t$ to be the $\sigma$-algebra generated by $\{X(s), s\ge t\}$. Let $f:\mathbb{Z}\times {\cal X}\rightarrow\R$. 
Let us compute the {\em forward difference} $\Delta^+f(t,X(t))$ with respect to the family $\{{\cal F}^-_t\}, t\ge 0,$ following Appendix \ref{Nelson}. 
We have that \begin{equation}\label{forgen}\Delta^+f(t,X(t))_{|_{X(t)=i}}=\E(f(t+1,X(t+1))-f(t,X(t))|X(t)=i)=\sum_jf(t+1,j)p_{ij}(t)-f(t,i).
\end{equation} 
Henceforth, we shall denote by $f_t$ and $\Delta^+f_t$ the column vectors with $i$-th component $f(t,i)$ and $\Delta^+f(t,X(t))_{|_{X(t)=i}}$, respectively. We can then rewrite (\ref{forgen}) in the compact form
\beq \label{generatorchain}
\Delta^+f_t=P(t)f_{t+1}-f_t.
\eeq
Consider now the vector space
$${\cal K}=\{f:\mathbb{Z}\times {\cal X}\rightarrow\R\,|\,\exists\,  t_0,t_1, t_0\leq t_1 \mbox{ s. t. } f(t,i)=0, \forall i,t\notin [t_0,\,t_1]\},$$ 
namely the set of functions with finite support. 
For $f,g\in{\cal K}$, we define the semi-definite {\em space-time inner product} as
\beq \label{sptimeip} \langle f,g\rangle_\pi=\sum_{t=-\infty}^\infty\langle f_t,g_t\rangle_{\pi_t}=\sum_{t=-\infty}^\infty f_t^\dag D_{\pi_t} g_t,=\sum_{t=-\infty}^{\infty} \E(f(t,X(t))\,g(t,X(t))),\eeq
where $\pi\sim\{\pi_t, t\in\Z\}$ denotes the family of the Markov chain distributions. We then have the following Corollary to the ``integration by parts" formula of Theorem \ref{intbyparts}.
\begin{corollary} Let $f,g\in{\cal K}$. Then
\beq\label{stadj}\langle \Delta^+f,g\rangle_\pi=\langle f,\Delta^-g\rangle_\pi\eeq
\end{corollary}
\proofn
By Theorem \ref{intbyparts},
\beqa
&&\nonumber\sum_{t=-\infty}^{\infty} \E(\Delta^+f(t,X(t))\,g(t,X(t)))=\sum_{t=-\infty}^{\infty}\E\left(f(t+1,X(t+1))\Delta^- g(t+1,X(t+1))\right)\\ &&=\sum_{s=-\infty}^{\infty}\langle f_{s},\Delta^- g_{s}\rangle_{\pi_{s}}\label{deltadj}
\eeqa
since the boundary terms are zero. \qed
\noindent
In view of relation \eqref{stadj}, we call $\Delta^-$ a $\langle\cdot,\cdot\rangle_\pi$-adjoint of $\Delta^+.$
Hence, the two conditional differences are {\em adjoint} with respect to the semi-definite {\em space-time inner product}. On the other hand, by using (\ref{generatorchain}), we get 
\beqa &&\nonumber\sum_{t=-\infty}^{\infty} \E(\Delta^+f(t,X(t))\,g(t,X(t)))=\sum_{t=-\infty}^{\infty}\sum_i\Delta^+f(t,X(t))_{|_{X(t)=i}}g(t,i)\pi_i(t)=\sum_{t=-\infty}^{\infty}\langle \Delta^+f_t,g_t\rangle_{\pi_t}\\
&&\nonumber=\sum_{t=-\infty}^{\infty}\langle P(t)f_{t+1}-f_t,g_t\rangle_{\pi_t}=\sum_{t=-\infty}^{\infty}f_{t+1}^\dag P^\dag(t)D_{\pi_t}g_t-\sum_{t=-\infty}^{\infty}f_t^\dag D_{\pi_t}g_t\\
&&\nonumber=\sum_{t=-\infty}^{\infty}f_{t+1}^\dag D_{\pi_{t+1}}D^{-1}_{\pi_{t+1}} P^\dag(t)D_{\pi_{t}}g_t-\sum_{t=-\infty}^{\infty}f_{t+1}^\dag D_{\pi_{t+1}}g_{t+1}\\
&&\label{quasiadj}=\sum_{t=-\infty}^{\infty}\langle f_{t+1}, D^{-1}_{\pi_{t+1}} P^\dag(t)D_{\pi_{t}}g_t-g_{t+1}\rangle_{\pi_{t+1}}\nonumber
\eeqa
Let $\pi_t(i)>0$ for all $t,i.$ In this case, (\ref{sptimeip})  is an inner product and the corresponding  adjoint is unique. By comparison with \eqref{deltadj}, we conclude that
\beq\label{forbacktrans}
\Delta^- g_{t+1}=D^{-1}_{\pi_{t+1}} P^\dag(t)D_{\pi_{t}}g_t-g_{t+1}.
\eeq
More explicitly, defining the matrices 
\beq\label{Q} Q(t)=D^{-1}_{\pi_{t+1}} P^\dag(t)D_{\pi_{t}},
\eeq
relation \eqref{forbacktrans} reads component-wise
\beqa\Delta^-g(t+1,X(t+1))_{|_{X(t+1)=j}}&=&\E(g(t,X(t)-g(t+1,X(t+1))|X(t+1)=j)\\&=&\sum_ig(t,i)q_{ji}(t)-g(t+1,j).
\eeqa 
Hence, $Q(t)$ is simply the matrix of the reverse-time transition probabilities. Of course, $Q$ can be obtained immediately by requiring that the two-time probabilities generated by the forward and backward Markov chains are the same:
\beq\label{bayes}\Prob(X(t)=i,X(t+1)=j)=p_{ij}(t)\pi_t(i)=q_{ji}\pi_{t+1}(j).\eeq
This yields immediately
$$q_{ji}(t)=p_{ij}(t)\frac{\pi_t(i)}{\pi_{t+1}(j)},$$
which can be compactly rewritten in the form (\ref{Q}).

Two remarks are in order: (i) The backward transitions are  time-dependent even when the forward are not. (ii) When $\pi_{t+1}(j)=0$, $q_{ji}(t)$ may be defined arbitrarily to be any number between zero and one without actually affecting relation \eqref{bayes}, provided  it satisfies the normalization condition
$$\sum_iq_{ji}(t)=1.
$$
Notice then that \eqref{quasiadj} leads to the correct form of the time-reversal even if the distributions $\{\pi_t\}$ are only non-negative.  
The derivation of $Q$ using the $\Delta^-$ operator, albeit much longer, permits to see that the reverse time transition mechanism may be viewed as a space-time adjoint to the forward one with respect to the flow of probability distributions $\{\pi_t, t\in\Z\}$.  The space-time adjointness relation \eqref{stadj} for Markov chains admits an {equivalent}, compact formulation. 

\begin{proposition} The space-time adjointness relation \eqref{stadj} holds if and only if the two-time relation 
\beq\label{staradjoint} \langle P(t)x,y\rangle_{\pi_t} =\langle x, Q(t) y\rangle_{\pi_{t+1}},\quad x,y\in\R^n,\eeq
is satisfied at any $t$.
\end{proposition}

\proofn It is immediate to verify that the form of the time-reversal transition matrix \eqref{Q} implies \eqref{staradjoint}. On the other hand, if (\ref{staradjoint}) holds, we have
\beqa &&\sum_{t=-\infty}^{\infty}\langle f_{t+1}, D^{-1}_{\pi_{t+1}} P^\dag(t)D_{\pi_{t}}g_t-g_{t+1}\rangle_{\pi_{t+1}}=\sum_{t=-\infty}^{\infty}\langle f_{t+1}, Q(t)g_t-g_{t+1}\rangle_{\pi_{t+1}}\\
&&\nonumber=\sum_{t=-\infty}^{\infty}\langle P(t)f_{t+1},g_t\rangle_{\pi_t}-\sum_{t=-\infty}^{\infty}\langle f_{t+1},g_{t+1}\rangle_{\pi_{t+1}}\\
&&\nonumber=\sum_{t=-\infty}^{\infty}\langle P(t)f_{t+1},g_t\rangle_{\pi_t}-\sum_{t=-\infty}^{\infty}\langle f_{t},g_{t}\rangle_{\pi_{t}}\\&&\nonumber=\sum_{t=-\infty}^{\infty}\langle P(t)f_{t+1}-f_t,g_t\rangle_{\pi_t}.
\eeqa
\qed

\noindent Relation (\ref{staradjoint}) will serve as a useful guideline to derive the reverse-time transition mechanism for quantum channels in Section \ref{Qrev}, since in that setting we cannot generally relay on conditional probabilities as in \eqref{bayes}.

\subsection{Space-time harmonic processes}\label{martingale} 
>From here on, we only consider Markov chains $X=\{X(t),t \geq 0\}$ with values in ${\cal X}=\{1,2,\ldots,n\}$.
\begin{definition} A function $h:\N\times{\cal X}\rightarrow \R$ is called {\em space-time harmonic} on $[t_0,t_1]$ for the transition mechanism $\{P(t);t_0\le t\le t_1\}$ of a chain if, for every $t_0\le t\le t_1-1$ and all $i\in{\cal X}$, it satisfies the {\em backward equation}
\begin{equation}h(t,i)=\sum_{j}p_{ij}(t)h(t+1,j).
\end{equation}
\end{definition}
The concept of space-time harmonic function can be introduced also with respect to a reverse time mechanism. Indeed, let $q_{ji}(t), t\ge 0$ be the reverse-time transition probabilities of the Markov chain  $X=\{X(t); t\in\N\}$. Then $\theta$ is called reverse-time harmonic with respect to $q_{ji}(t), t\ge 0$ if it satisfies
\begin{equation}\label{rtspace-time}
\theta(t+1,j)=\sum_iq_{ji}(t)\theta(t,i), \quad \forall t\ge 0, \forall i,j\in{\cal X}.
\end{equation}

Space-time harmonic functions, a terminology due to Doob and motivated by diffusion processes, play a central role in constructing Schr\"{o}dinger bridges for Markov chains \cite{pavon-bridges}. They are closely related to a class of {\em martingales} that are  {\em instantaneous functions} of $X(t)$. Let, as before, ${\cal F}^-_t$ denote the $\sigma$-algebra induced by $\{X(s), s\le t\}$. 

\begin{proposition}\label{spacemart} Let $h$ be space-time harmonic on $[t_0,t_1]$ for the (transition mechanism of the) Markov chain $X=\{X(t); t\in\N\}$ with state space ${\cal X}$ and transition matrix $P(t)=\left(p_{ij}(t)\right)$. Define the stochastic process $Y=\{Y(t)=h(t,X(t)), t_0\le t\le t_1\}$. Then,  $Y$ is a martingale with respect to $\{{\cal F}^-_t, t_0\le t\le t_1\}$. 
\end{proposition}
The proof is a straightforward generalization of Bremaud \cite[p.179]{bremaud}.
By Jensen's inequality \cite{rudin}, we have the following way to generate submartingales from martingales.
\begin{proposition}\label{submart}{ Let $Y=\{Y(t), t\ge 0\}$ be a martingale with respect to the filtration $\{{\cal F}_t, t\ge 0\}$ induced by the past of the Markov chain $X=\{X(t), t\ge 0\}$. Let $\varphi$ be a convex function and define $Z(t):=\varphi(Y(t)), t\ge 0$. Then $Z$ is a submartingale with respect to $\{{\cal F}_t, t\ge 0\}$, namely
$$\E(Z(t+1)|{\cal F}_{t})\ge Z(t),\quad a.s.
$$}
\end{proposition}


\section{Time-reversal for quantum Markov channels}\label{Qrev}
Consider an $n$-level quantum system  with associated Hilbert space $\Hi$ isomorphic to ${\mathbb C}^n$. In its standard statistical description, the role of probability densities is played by density operators, namely by positive, unit-trace matrices $\rho\in{\cal D}(\Hi)$. The  role of real random variables is taken by Hermitian operators $X\in{\cal O} (\Hi)$ representing obervables. Expectations are computed via the trace functional, $\E_\rho(X)=\trace(\rho X),$ and the classical setting may be recovered considering all diagonal matrices. Any linear, Trace Preserving and Completely Positive (TPCP) dynamical map $\Ec^\dag$ acting on  density operators can be represented by a Kraus operator-sum \cite{kraus}, i.e.:  
$$\rho_{t+1}=\Ec^\dag(\rho_t)=\sum_j M_j\rho_t M^\dag_j,\quad \sum_j M_j^\dag M_j = I.$$
Following a quite standard quantum information terminology, we refer to linear, completely-positive trace-non-increasing Kraus maps as {\em quantum operations}. The adjoint action of a quantum operation with respect to the Hilbert-Schmidt inner product, i.e. the expectation, is immediately found: 
$$\trace(X\Ec^\dag(\rho))=\trace(X\sum_jM_j\rho M_j^\dag)=\trace(\sum_jM_j^\dag X M_j\rho)=\trace(\Ec(X)\rho).$$
For observables, the dual dynamics is thus given by the {\em identity-preserving} quantum operation \beq\label{IP}\Ec(X)=\sum_jM_j^\dag X M_j.\eeq In the remaining of the paper, we consider the discrete-time quantum Markov evolutions associated to an initial density matrix $\rho_0$ and a sequence of TPCP maps $\{\Ec^\dag_t\}_{t\geq0}.$

In order to find the time-reversal of a given Markovian evolution, rewrite the probability-weighted inner product of the classical case \eqref{scalar} as
$\langle x,y\rangle_\pi=\trace(D_x D_\pi D_y).$ Notice that, if we simply drop commutativity, for two observables $X,Y$ and a density matrix $\rho$, we would obtain $\langle X,Y\rangle_\rho=\trace(X\rho Y).$ This functional is not satisfactory to our scopes, since in general it is neither real nor symmetric, i.e. $\trace(Y\rho X)\neq \trace(X\rho Y).$ It is then convenient to rewrite \eqref{scalar}, by using the fact that all matrices commute, in the symmetrized form:
$$\langle x,y\rangle_\pi=\trace(D_x^{\frac{1}{2}} D_\pi^{\frac{1}{2}} D_y D_\pi^{\frac{1}{2}}D_x^{\frac{1}{2}}).$$
We shall show that this form of the inner product leads to the correct reverse-time quantum Markov operation. Allowing for a general density operator $\rho$ and observables $X,Y$, we thus define:
$$\langle X,Y\rangle_\rho=\trace(X^\um\rho^\um Y\rho^\um X^\um).$$
This is a symmetric, real, semi-definite sesquilinear form on Hermitian operators.

By analogy with the classical case, we then define the quantum operation $\Rc_{\Ec,\rho_t}$ as the space-time $\{\rho_t\}$-adjoint of a quantum operation $\Ec$ using the quantum version of \eqref{staradjoint}:
$$\langle \Ec(X),Y \rangle_{\rho_t}=\langle X,\Rc_{\Ec,\rho_t}(Y) \rangle_{\rho_{t+1}}.$$
Let us assume for now that $\rho_{t+1}$ is full-rank. An explicit Kraus representation is then obtained as follows:
\beqan \langle \Ec(X),Y \rangle_{\rho_t}&=& \sum_j \trace(M_j^\dag X M_j\rho_t^\um Y \rho_t^\um)\\
&=& \sum_j \trace( X \rho^\um_{t+1}\rho^{-\um}_{t+1}M_j\rho_t^\um Y \rho_t^\um M_j^\dag\rho^{-\um}_{t+1}\rho^\um_{t+1})\\
&=& \sum_j \trace(X \rho^\um_{t+1}R^\dag_j(\Ec,\rho_t)Y R_j(\Ec,\rho_t)\rho^\um_{t+1})\\
&=& \langle X, \Rc_{\Ec,\rho_t}(Y) \rangle_{\rho_{t+1}},
\eeqan
where $\Rc_{\Ec,\rho_t}$ admits an operator-sum representation with Kraus operators \beq \label{qreverse} R_j(\Ec,\rho_t)=\rho^{-\um}_{t+1}M_j\rho_t^\um.\eeq
Notice that the second equality is non-trivial in the case when $\rho_{t+1}$ is not full-rank and inverses are replaced by the Moore-Penrose pseudoinverse (the latter replacement will be tacitly assumed in the rest of the paper). For any matrix $M$, the {\em support} of $M$, denoted $\supp(M)$, is the orthogonal complement of $\ker(M)$. The following lemma ensures that the same derivation applies to the general case.

\begin{lemma}\label{lemma} Let $\rho_{t+1}=\sum_jM_j\rho_t M_j^\dag$. Let $\Pi_{\rho_{t+1}}$ denote the orthogonal projection onto the support of $\rho_{t+1}$. Then, for any normal matrix $Y$:
$$\Pi_{\rho_{t+1}}\left(\sum_jM_j\rho_t^\um Y \rho_t^\um M_j^\dag\right)\Pi_{\rho_{t+1}}=\sum_jM_j\rho_t^\um Y \rho_t^\um M_j^\dag.$$ \end{lemma}

\proofn   
If we consider a spectral representation for $\rho_t^\um=\sum_k\sqrt{p_k}|\alpha_k\rangle\langle\alpha_k|,$ we have that, for any $|y\rangle$:
\beqan\sum_jM_j\rho_t^\um |y\rangle\langle y | \rho_t^\um M_j^\dag&=&\sum_jM_j\sum_{k,l} \sqrt{p_k p_l} |\alpha_k\rangle\langle\alpha_k|y\rangle \langle y|\alpha_l\rangle\langle\alpha_l|M_j^\dag\\&=&\sum_j\sum_{k,l} \sqrt{p_k p_l} y_{k,l} M_J|\alpha_k\rangle\langle\alpha_l|M_J^\dag,\eeqan
where $y_{k,l}=\langle\alpha_k|y\rangle \langle y|\alpha_l\rangle.$ Since $\rho_{t+1}=\sum_{j,k} p_k M_j|\alpha_k\rangle\langle\alpha_k |M^\dag_j$ it must be $\Pi_{\rho_{t+1}}^\perp  M_J|\alpha_k\rangle = 0$ for all $j,k.$  Hence
$$\Pi_{\rho_{t+1}}^\perp\sum_jM_j\rho_t^\um |y\rangle\langle y | \rho_t^\um M_j^\dag\Pi_{\rho_{t+1}}^\perp = 0.$$
and the statement holds for rank one $Y=|y\rangle \langle y |$. By linearity it extends to any normal matrix.
\qed

It is now natural to define a transformation between Kraus operators. Let $\Ec^\dag$ be a quantum operation represented by Kraus operators $\{F_k\}.$ For any $\rho,$ define the map ${\cal T}_{\rho}$ from quantum operations to quantum operations  
\beq {\cal T}_{\rho}:\Ec^\dag\mapsto {\cal T}_{\rho}(\Ec^\dag),\eeq
where ${\cal T}_{\rho}(\Ec^\dag)$ has Kraus operators $\{\rho^\um F^\dag_k (\Ec(\rho))^{-\um} \}.$  The results of  \cite{barnum-reversing} show that the action of ${\cal T}_{\rho}$ is independent of the particular Kraus representation of $\Ec^\dag$.
With this definition, we have that $${\cal T}_{\rho_t}(\Ec^\dag)=\Rc^\dag_{\Ec,\rho_t}.$$

We are now in a position to prove the main result of this section, which establishes the role of $\Rc_{\Ec,\rho_t}(\cdot)$ as the quantum time-reversal of the TPCP map $\Ec^\dag.$ Augmenting a Kraus map $\Ec$ with Kraus operators $\{ M_k \}_{k=1,\ldots, m}$ to a TPCP map means adding a finite number $p$ of Kraus operators $\{ M_k \}_{k=m+1,\ldots, m+p}$ so that $\sum_k M^\dag_k M_k = I.$

\begin{theorem} [Time Reversal of TPCP maps] Let  $\Ec^\dag$ be a TPCP map. If $\rho_{t+1}=\Ec^\dag(\rho_t),$ then for any $\rho_t\in\mathfrak{D}({\cal H}),$ $\Rc^\dag_{\Ec,\rho_t}={\cal T}_{\rho_t}(\Ec^\dag)$ defined as in \eqref{qreverse} is the {\em time-reversal} of $\Ec^\dag$ for $\rho_t$, that is, it satisfies both: 
\beq\label{revprop}\rho_t=\Rc^\dag_{\Ec,\rho_t}(\rho_{t+1})\eeq
 and:
\beq\label{consistency} {\cal T}_{\rho_{t+1}}(\Rc^\dag_{\Ec,\rho_t})(\sigma_t)=\Ec^\dag(\sigma_t),\eeq
for all $\sigma_t\in{\frak D}(\Hi)$ such that {\em $\supp(\sigma_t)\subseteq \supp(\rho_{t}).$ } 
Morover, it can be augmented to be TPCP without affecting property \eqref{revprop}-\eqref{consistency}.
\end{theorem}

\proofn By direct calculation:
\beqan 
\Rc^\dag_{\Ec,\rho_t}(\rho_{t+1})&=& \sum_j \rho_t^\um M_j^\dag \rho^{-\um}_{t+1} \rho_{t+1} \rho^{-\um}_{t+1}M_j\rho_t^\um\\
&=& \sum_j \rho_t^\um M_j^\dag \Pi_{\rho_{t+1}}M_j\rho_t^\um.
\eeqan
If $\rho_{t+1}$ is full-rank we are done, since $\sum_jM_j^\dag M_j=I.$ If this not the case, consider an orthonormal basis $\{| \alpha_k\rangle\}$ for $\ker(\rho_{t+1}).$
Observe that:
$$ 0=\langle\alpha_k|\rho_{t+1}|\alpha_k\rangle=\langle\alpha_k|\sum_jM_j\rho_tM_j^\dag|\alpha_k\rangle,$$ 
which implies $M_j^\dag|\alpha_k\rangle\in\ker(\rho_t),\,\forall j,k.$
Now decompose the identity operator as follows:
$$I=\sum_jM_j^\dag M_j=\sum_jM_j^\dag\Pi_{\rho_{t+1}} M_j+\sum_jM_j^\dag\sum_k |\alpha_k\rangle\langle\alpha_k|  M_j,$$
and multiply on both sides by $\Pi_{\rho_t}.$ Since $M_j^\dag|\alpha_k\rangle\in\ker(\rho_t),$ we obtain:
$$\Pi_{\rho_t}=\sum_j \Pi_{\rho_t} M_j^\dag \Pi_{\rho_{t+1}} M_j \Pi_{\rho_t}.$$
Hence $\sum_j \rho_t^\um M_j^\dag \Pi_{\rho_{t+1}}M_j\rho_t^\um=\rho_t.$

\noindent In order to prove \eqref{consistency}, recall that $\Rc^\dag_{\Ec,\rho_t}$ admits Kraus operators $R_j(\Ec,\rho_t)=\rho_t^\um M^\dag_k \rho_{t+1}^{-\um}.$ If we explicitly compute ${\cal T}_{\rho_{t+1}}(\Rc^\dag_{\Ec,\rho_t})$ we get a quantum operation with Kraus operators $$\rho_{t+1}^{\um}R^\dag_j(\Ec,\rho_t)\rho^{-\um}_{t}=\rho_{t+1}^{\um} (\rho^{-\um}_{t+1} M_k  \rho_{t}^{\um} )\rho^{-\um}_{t}=\Pi_{\rho_{t+1}} M_k  \Pi_{\rho_t}.$$
Hence, if $\Pi_{\rho_t}\sigma_t\Pi_{\rho_t}=\sigma_t$ by Lemma \ref{lemma} we get:
$${\cal T}_{\rho_{t+1}}(\Rc^\dag_{\Ec,\rho_t})(\sigma_t)=\sum_k\Pi_{\rho_{t+1}} M_k  \Pi_{\rho_t}\sigma_t\Pi_{\rho_{t}} M^\dag_k  \Pi_{\rho_t+1}=\sum_kM_k  \Pi_{\rho_t}\sigma_t\Pi_{\rho_{t}} M^\dag_k=\Ec^\dag(\sigma_t).$$
In general, $\Rc^\dag_{\Ec,\rho_t}$ is trace-non-increasing, since:
\beqan \sum_j R_j^\dag(\Ec,\rho_t)R_j(\Ec,\rho_t)&=&\rho^{-\um}_{t+1}\sum_jM_j\rho_t M_j^\dag\rho^{-\um}_{t+1} \\
&=& \Pi_{\rho_{t+1}}.\eeqan
If $\rho_{t+1}$ is full rank, then $\Rc^\dag_{\Ec,\rho_t}$ is trace preserving.
If this is not the case, we can always ``augment'' $\Rc^\dag_{\Ec,\rho_t}$ with some additional Kraus operators $\tilde R_j$ that satisfy:
$$\sum_j \tilde R_j^\dag \tilde R_j= I - \Pi_{\rho_{t+1}},\quad \tilde R_j\Pi_{\rho_{t+1}}\tilde R_j^\dag=0,$$
 so that the augmented $\Rc^\dag_{\Ec,\rho_t}$ is trace-preserving and does act as the time reversal on $\rho_{t+1}$. To do so, it suffices for example to consider an orthonormal basis $\{ |\beta_j\rangle\}$ for $\ker(\rho_{t+1}),$ and define $\tilde R_j=|\beta_j\rangle\langle \beta_j |.$
\qed

\noindent {\bf Remark:} Property \eqref{consistency} ensure us that among all quantum operations mapping $\rho_{t+1}$  back to $\rho_t,$ $\Rc^\dag_{\Ec,\rho_t}$ is the natural {\em time-reversal} of $\Ec^\dag$ with respect to $\rho_t$. In fact, notice that if $\rho_t$ is full rank, \eqref{consistency} implies that ${\cal T}_{\rho_{t+1}}\circ{\cal T}_{\rho_{t}}$ is the the identity map on quantum operations. That is, as one would expect, the time reversal of the time-reversal is the original forward map. While this may seem obvious, notice that property \eqref{revprop} alone is satisfied by any quantum operation of the form $$\tilde\Rc^\dag={\cal T}_{\rho_t}(\Fc^\dag),$$ with $\Fc^\dag$ {\em any} TPCP map.

\section{Other approaches to Quantum Time-Reversal}\label{other}

Reversibility issues for quantum operations have been related to QEC from its beginning to the most recent approaches, see e.g. \cite{NC-rev,knill-qec,kribs-OEC,viola-IPS}. While studying quantum error correction problems, the same $\Rc^\dag_{\Ec,\rho}(\cdot)$ has been suggested by Barnum and Knill as a near-optimal correction operator \cite{barnum-reversing}. 
They introduce the explicit form of $\Rc^\dag_{\Ec,\rho}(\cdot)$ by analogy with the error correction operation for subspace codes. The correction operation for a TPCP map with Kraus representation $\{ M_k\}$ features Kraus operators $\{  \Pi_CM_k^\dag /\sqrt{p_k}\}$. Here $ \Pi_C$ is the orthogonal projection on the subspace code, and $M_k\Pi_C=\sqrt{p_k}V_k$ has to hold for some probabilities $\{p_k\}$ and isometries $V_k$ on orthogonal subspaces.
In their setting, $\Ec^\dag(\rho)$ is assumed to be full-rank, and represents the output of a channel $\Ec^\dag$ with input state $\rho=\sum_j p_j\rho_j,$ a statistical mixture of some quantum codewords of interest to be recovered. 
$\Rc^\dag_{\Ec,\rho}(\cdot)$ is then shown to satisfy $\Rc^\dag_{\Ec,\rho}(\Ec^\dag(\rho))=\rho.$ It is also proven there that the reversal is independent of the particular Kraus representation of $\Ec^\dag$, a fact that we used in the previous section to introduce ${\cal T}_{\rho}$.

Another approach, which is strictly related to our work in \cite{pavon-bridges}, is based on the fact that density operators $\rho_t,\rho_{t+1}$ can be interpreted as classical probability distributions over their spectral families of orthogonal projectors. Each set of orthogonal projectors is a complete family of {\em commuting quantum events} that generates an abelian algebra. We can thus study the transition probabilities between the elementary events of the abelian algebras generated by $\rho_t,\rho_{t+1}$ following the analogy with the classical case.
In fact, if we write $\rho_t=\sum_ip_i\Pi_{t,i},$ $\rho_{t+1}=\sum_jq_j\Pi_{t+1,j},$ we have that the probability of measuring $\Pi_{j,t+1}$ after $\Pi_{i,t}$ has been measured at time $t$ and $\Ec^\dag$ acted on the system, is given by: 
\beqa \Prob(\Pi_{i,t},\Pi_{j,t+1})&=&\trace\left(\Pi_{j,t+1}\left(\sum_kM_k \Pi_{i,t}\rho_t\Pi_{i,t} M_k^\dag\right) \Pi_{j,t+1}\right),\nonumber\\
&=& \trace\Big(\Pi_{j,t+1}(\sum_kM_k \Pi_{i,t} M_k^\dag) \Pi_{j,t+1}\Big)p_i,\label{forwardp}
\eeqa
obtaining an analogous of the transition probabilities in the classical case. The reverse-time $\Rc^\dag_{\Ec,\rho_t}$ is then required to be such that:
\beq\label{revprob}\Prob(\Pi_{i,t},\Pi_{j,t+1})=\trace\Big(\Pi_{j,t}\sum_kR_k(\Ec,\rho_t) \Pi_{i,t+1} R_k^\dag(\Ec,\rho_t) \Pi_{j,t}\Big)q_j.\eeq
From \eqref{forwardp}, using the cyclic property of the trace and the fact that $\rho_{t+1}$ and $\Pi_{j,t+1}$ commute for all $j$, we get:
\beqa \Prob(\Pi_{i,t},\Pi_{j,t+1})&=&\sum_k\trace(\Pi_{j,t+1}M_k \Pi_{i,t}\rho_t\Pi_{i,t} M_k^\dag \Pi_{j,t+1}),\nonumber\\
&=&\sum_k  \trace\Big(\Pi_{i,t}\rho_t^\um M^\dag_k\rho_{t+1}^{-\um} \Pi_{j,t+1} \rho_{t+1}\Pi_{j,t+1}\rho_{t+1}^{-\um}M_k \rho_t^\um\Pi_{i,t}\Big),\nonumber\\
&=& \trace\Big(\Pi_{i,t}\sum_k (\rho_t^\um M^\dag_k\rho_{t+1}^{-\um}) \Pi_{j,t+1} (\rho_{t+1}^{-\um}M_k \rho_t^\um)\Pi_{i,t}\Big)q_j.\nonumber
\eeqa
This shows that $\Rc^\dag_{\Ec,\rho_t}$ we derived before satisfies \eqref{revprob}. This property has been used in \cite{pavon-bridges} to intuitively derive the form of $\Rc^\dag_{\Ec,\rho_t}$, where the time-reversal has been proved to be a key ingredient to solve {\em maximum entropy problems}
on quantum path spaces. In Appendix \ref{pathsection} we outline some connections between the material presented in this paper and Theorem 6.5 in \cite{pavon-bridges}.

Yet another possibility to approach the time-reversal is offered by the interpretation of a TPCP map in Kraus form as a non-selective generalized measurement (see e.g. \cite{nielsen-chuang}). 
General, indirect quantum measurements cause the initial state $\rho_t$ to ``collapse'' onto one of the conditional density operators: $$\rho_k=\frac{1}{\trace(M_k^\dag M_k \rho_t )}M_k \rho_t M_k^\dag,$$
with relative probabilities $p(k)=\trace(M_k^\dag M_k \rho_t ),$ for some $\sum_k M^\dag_k M_k =I.$ One can then think of  $\Ec(\rho_t)=\sum_k M_k (\rho_t) M_k^\dag$ as the state conditioned after a non-selective quantum measurement with outcomes labeled by $k,$ i.e. a measurement with unknown outcome. 
We can then look for a sort of reverse-time measurement process, namely a quantum operation that has the same reverse transition probabilities. Define as above $\rho_{t+1}=\Ec(\rho_{t})$. We look for a trace-preserving
$\Rc_{\Ec,\rho_t}(\cdot)=\sum_k R_k(\Ec,\rho_t) (\cdot) R_k^\dag(\Ec,\rho_t)$ such that:
$$\trace(M_k^\dag M_k \rho_t)=p(k)=\trace(R_k^\dag(\Ec,\rho_t) R_k(\Ec,\rho_t)\rho_{t+1}). $$
By using again the cyclic property of trace, we get:
$$p(k)=\trace(M_k^\dag M_k \rho_t)= \trace( \rho^{-\um}_{t+1}M_k \rho_t M_k^\dag\rho^{-\um}_{t+1}\rho_{t+1}),$$
which suggests to the same form of the time reversal map we obtained in the previous section, namely $R_k(\Ec,\rho_t)= \rho_t^{\um}M^\dag_k\rho_{t+1}^{-\um}. $ This can be seen as a simplification of the three-time derivation proposed in \cite{crooks} for quantum operations at the equilibrium.

\section{Quantum space-time harmonic processes}\label{Qsth}

While in the framework of quantum probability rigorous extensions of conditional expectations and martingale processes are available for quite some time \cite{takesaki,accardi-qsp,parthasaraty,kummerer}, we show here that quantum analogues of the results of Section \ref{classicalresults} below can be derived avoiding most of the related technical machinery. This can be accomplished by introducing a quantum version of space-time harmonic functions.
Consider a reference quantum Markov evolution on a finite time interval, generated by an initial density matrix $\rho_0$ and a sequence of TPCP maps $\{\Ec^\dag_t\}_{t\in[0, T-1]}.$
\begin{definition} [Quantum space-time harmonic process] \label{qharmonic} A sequence of Hermitian operators $\{Y_t\}_{t\in[0, T-1]}$ is said to be space-time harmonic with respect to the family of identity-preserving maps $\{\Ec_t\}_{t\in[0, T-1]}$ if:
\beq \label{qsth} Y_{t}=\Ec_t(Y_{t+1}). \eeq
\end{definition}
In analogy with the classical case, $\{Y_t\}_{t\in[0, T-1]}$ is said to be {\em space-time harmonic in reverse-time} with respect to the family $\{\Rc_{\Ec_T,\rho_t}\}$ if:
\beq \label{qsthr} Y_{t+1}=\Rc_{\Ec_t,\rho_t}(Y_t). \eeq
The sequence is called {\em space time subharmonic} if $Y_{t}\le \Ec_t(Y_{t+1})$, where we are referring to the natural partial order between Hermitian matrices, see also Section \ref{jensen}. Similarly in reverse time. 

In the classical case, space time harmonic functions generate changes of measure through multiplicative functional transformations of the transition mechanism. A similar fact holds in the quantum case. Let $Y_t$ be space time harmonic for $\Ec_t \sim \{E_k(t)^\dag\}$ and let $N_t$ be any choice of operator such that $Y_t=N_t N_t^\dag.$ Assume for simplicity $Y_t$ to be full-rank at any $t$. Then $\Fc_t \sim \{ N_t^{-1} E_k(t)^\dag N_{t+1}\}$ is an identity-preserving quantum operation. In fact, by using \eqref{qsth}, we have:
$$\Fc_t(I)=\sum_kN_t^{-1} E_k(t)^\dag N_{t+1}N_{t+1}^\dag E_k(t)N_t^{-\dag}=I.$$
Thus its adjoint is a TPCP map. An analogous result holds for reverse time evolution.
The following result is the quantum counterpart of  (\ref{constexpmart}) concerning properties of expectation of (sub)martingales.
\begin{proposition} Let $\{Y_t\}_{t\in[0, T-1]}$ be space-time harmonic and let $\{Z_t\}_{t\in[0, T-1]}$ be space-time subharmonic with respect to the reference evolution. Then, for all $t\in[0, T-1]$:
\beq\label{constantexp}\E_{\rho_0}(Y_0)=\E_{\rho_{t}}(Y_{t}),\quad \E_{\rho_{t}}(Z_{t})\le\E_{\rho_{t+1}}(Z_{t+1}).\eeq
\end{proposition}
\proofn
Using \eqref{qsth}, we have
$$\tr(\rho_{t+1}Y_{t+1})=\tr(\Ec_t^\dag(\rho_{t})Y_{t+1})=\tr(\rho_{t}\Ec_t(Y_{t+1}))=\tr(\rho_{t}Y_{t}).$$
We then get \eqref{constantexp} by iterating the above calculation.  Similarly,
$$\tr(\rho_{t+1}Z_{t+1})=\tr(\Ec_t^\dag(\rho_{t})Z_{t+1})=\tr(\rho_{t}\Ec_t(Z_{t+1}))\ge\tr(\rho_{t}Z_{t}).$$
\qed

We are now ready for the quantum counterpart of Proposition \ref{submart}.

\begin{proposition}\label{operatorsubmartingale}  Let $Y_t$ be a space-time harmonic process with respect to $\{\Ec_t\}_{t\geq 0},$ with eigenvalues $\lambda_{t,i}\in{\cal I}\subset \R$ at all times, and $f:{\cal I}\rightarrow \R$ be operator convex. Then $Z_t:=f(Y_t)$ is space-time subharmonic.
\end{proposition} 
\proofn
By Definition \ref{qharmonic} and Theorem \ref{operatorjensen}, we obtain:
\beqan Z_t=f(Y_{t})&=& f(\Ec_t(Y_{t+1}))\leq \Ec_t(f(Y_{t+1}))=\Ec_t(Z_{t+1}).
\eeqan\qed
\noindent 

\section{Application to thermodynamics}
\subsection{Classical results: Space-time harmonic functions and a strong form of the H-theorem}\label{classicalresults}

As first observed by Doob \cite{doob-markov}, the ratio of two solutions of a forward equation (\ref{piev}) yields a space-time harmonic function for the reverse-time transition mechanism. Indeed, let $\{\pi_t, t\ge 0\}$ and $\{p_t, t\ge 0\}$ both satisfy equation (\ref{piev}), with $\pi_t(i)>0, \forall i,  t_0\le t\le t_1$. Let $q^\pi_{ji}(t)$ denote the reverse-time transition probabilities corresponding to the initial condition $\pi(0)$. Define the function
\beq\theta(t,i):=\frac{p_t(i)}{\pi_t(i)}.\label{rate}
\eeq
Then, using (\ref{forbacktrans}) and (\ref{piev}), we get
\begin{equation}\label{rtspacetime}\sum_iq^\pi_{ji}(t)\theta(t,i)=\sum_i\frac{\pi_t(i)}{\pi_{t+1}(j)}\pi_{ij}(t)\frac{p_t(i)}{\pi_t(i)}=\theta(t+1,j),
\end{equation}
namely $\theta$ is space-time harmonic on $[t_0,t_1]$ with respect to the reverse-time  transition mechanism $q^\pi_{ji}(t)$. In view of Proposition \ref{spacemart}, it follows that the process $Y(t):=\theta(t,X(t))$ is a reverse-time martingale with respect to the ``future" filtrations $\{{\cal G}_t=\sigma(X(t),X(t+1),\ldots), t_0\le t\le t_1\}$.

\begin{theorem}\label{strong2law} Under the above assumptions, the stochastic process
$Z(t):=-\log Y(t):=-\log \theta(t,X(t))$ is a submartingale with respect to ${\cal G}_t, t_0\le t\le t_1$ in the reverse time direction.
\end{theorem}
\proofn Observe that $-\log$ is a convex function and invoke Proposition \ref{submart}.
\qed
\noindent
We now show that Theorem \ref{strong2law} implies a local form of the second law. Indeed, consider a time-homogenous chain with forward transition matrix $P$. Let $\bar{\pi}$ be a stationary distribution for the chain, namely $P^\dag\bar\pi=\bar\pi$.
Let $\pi_t$ be another solution of the corresponding forward equation (\ref{piev}). Assume $\pi_t(i)>0, \forall i,  \forall t\ge t_0$. Then, as observed before,
$$\bar{\theta}(t,i)=\frac{\bar{\pi}(i)}{\pi_t(i)}, t\ge t_0,
$$
is space-time harmonic on $t\ge t_0$ with respect to the reverse-time  transition mechanism $q^\pi_{ji}(t)$. By Theorem \ref{strong2law}, the process $\bar{Z}(t):=-\log \bar{\theta}(t,X(t))$ is a submartingale with respect to ${\cal G}_t, t\ge t_0$ (conditionally increasing) in the reverse time direction. Hence, we have the following strong form of the second law.
\begin{theorem} \label{classstronlaw}Under the above assumptions, we have
\beq \label{STRONGLAW}\E\left(-\log\bar{\theta}(t,X(t))\mid{\cal G}_{t+1}\right)\ge-\log\bar{\theta}(t+1,X(t+1)), \quad {\rm a.s.}.
\eeq
\end{theorem}
This stronger form of the second law was apparently first presented for diffusion processes in \cite{pavon}. We finally observe that the usual second law can be obtained as a consequence of this result (Corollary \ref{weakseclaw} below). 
Let us first recall the definition of relative entropy. Let $p$ and $q$ be probability distributions on a finite or countably infinite set. We say that the {\em support} of $p$ is contained in the support of $q$ if $q_i=0\Rightarrow p_i=0$ and write $\supp (p)\subseteq \supp (q)$. The  {\em Relative Entropy}  or {\em Information Divergence} or {\em Kullback-Leibler Index} of $q$ from $p$ is defined to be
\begin{equation}\label{KLdist}\D(p\|q)=\left\{\begin{array}{ll} \sum_ip(i)\log\frac{p(i)}{q(i)}, & \supp (p)\subseteq \supp (q),\\
+\infty , & \supp (p)\not\subseteq \supp (q).\end{array}\right.,
\end{equation} 
where, by definition, $0\cdot\log 0=0$.
\begin{corollary} \label{weakseclaw}$\D(\pi_t\|\bar{\pi})$ is nonincreasing.
\end{corollary}
\proofn
We take expectations in (\ref{STRONGLAW}). By the iterated conditioning property and observing that 
$$\E_{\pi(0)}(-\log \bar{\theta}(t,X(t))=\sum_i\log\frac{\pi_t(i)}{\bar{\pi}(i)}\pi_t(i)=\D(\pi_t\|\bar{\pi}),
$$
we get 
$$\D(\pi_{t+1}\|\bar{\pi})\le \D(\pi_t\|\bar{\pi}).
$$
\qed

\subsection{Relative entropies and a quantum H-theorem}\label{QHtheorem}

The usual definition of quantum relative entropy is due to Umegaki \cite{Ume}.
Given two density matrices $\rho,\sigma,$ the {\em quantum relative entropy} is defined as:
\begin{equation}\label{umegakirelentr}\D_U(\rho\|\sigma)=\left\{\begin{array}{ll} \tr(\rho(\log \rho -\log \sigma)), & \supp (\rho)\subseteq \supp (\sigma),\\
+\infty , & \supp (\rho)\not\subseteq  \supp (\sigma)\end{array}\right.,
\end{equation} 
As in the classical case, quantum relative entropy has the
property of a pseudo-distance (see e.g. \cite{nielsen-chuang, vedral-relativeentropy}).
Moreover, it has been proven by Petz that it is the only functional in a class of quasi-entropies having a certain conditional expectation property \cite{Petz-charofrelent}.

Nonetheless, here we show how a different quantum extension of classical relative entropy is natural from the viewpoint of space-time harmonic processes and the dynamical structure of Markovian evolutions.
In the classical case, we have that Kullback-Libler relative entropy between two probability densities $p_t,\pi_t$ can be obtained as: \beq\label{kl} \E_{\pi_t}\big(\tilde\theta(t,X(t))\log(\tilde\theta(t,X(t))\big)= \sum_i \pi_t(i) \frac{p_t(i)}{\pi_t(i)}  \log\left(\frac{p_t(i)}{\pi_t(i)}\right),\eeq
where $\tilde\theta(t,i)=\frac{p_t(i)}{\pi_t(i)}$ is space-time harmonic in reverse time if $p_t,\pi_t$ evolve with the same forward transition mechanism, see (\ref{rate})-(\ref{rtspacetime}). 

\noindent
We now introduce a class of space-time harmonic quantum processes that are the analogue of those in \eqref{rate}. Consider two quantum Markov evolutions, corresponding to different initial conditions $\rho_0\neq\sigma_0$, but with same family of trace-preserving quantum operations $\{\Ec^\dag_t\}$. Define the observable \beq\label{qsthrate} Y_t=\sigma_t^{-\um}\rho_t\sigma_t^{-\um}.\eeq
We thus have that:
\begin{equation}\label{rtspacetimeq}\Rc_{\Ec,\sigma_t}(Y_t)=\sum_k \sigma_{t+1}^{-\um} M_k \sigma_t^{\um} \sigma_t^{-\um}\rho_t\sigma_t^{-\um}\sigma_t^{\um}M_k^\dag \sigma_{t+1}^{-\um}=Y_{t+1}.
\end{equation}
This shows that $Y_t$ evolves in the forward direction with the backward transition mechanism of $\sigma_t,$ which makes it quantum {\em space-time harmonic in reverse time} with respect to the transition of $\sigma_t$.
In view of \eqref{kl} and \eqref{qsthrate}, the natural definition of relative entropy in our setting is thus the Belavkin-Staszewski's relative entropy \cite{B-S}: \beq\label{BSrelent} \D_{BS}(\rho||\sigma)= 
\trace\left(\sigma\left(\sigma^{-\um}\rho\sigma^{-\um}\right)\log\left(\sigma^{-\um}\rho\sigma^{-\um}\right)\right),\eeq
where, as usual, $0\log 0=0.$ As for the Umegaki's version, it enjoys the properties of a pseudo-distance: It is non negative and equal to zero if and only if $\rho=\sigma.$  
The proof is immediate by using \eqref{expectationjensen}. In addition to this, it is clearly consistent with the classical relative entropy, which is recovered by considering commuting matrices,  and with the von Neumann entropy, since:
$$\D_{BS}(\rho||I)= \trace(\rho\log(\rho)).$$ 
Another useful property has been proven by Hiai and Petz \cite{Hiai-P}:
\beq\label{H-P}\D_{BS}(\rho||\sigma)\geq\D_{U}(\rho||\sigma).\eeq
Hence, convergence in $\D_{BS}(\rho||\sigma)$ ensures convergence in $\D_{U}(\rho||\sigma).$ The Belavkin-Staszewski's relative entropy has also been shown to be the trace of 
Fuji-Kamei's operator entropy \cite{F-K}. As a consequence of the results of Section \ref{Qsth}, we have the following Corollary. 

\begin{corollary} \label{qsubm} Consider two quantum Markov evolutions associated to the initial conditions $\rho_0\neq\sigma_0$ and to the same family of TPCP maps $\{\Ec^\dag_t\}$. Suppose that $\rho_t,\sigma_t$ are invertible, for all $t$'s. Let $Y_t=\sigma_t^{-\um}\rho_t\sigma_t^{-\um}$ and let $Z_t:=g(Y_t),$ with $g(x)=x\log(x)$. Then $Z_t$ is a reverse time, space-time subharmonic process with respect to the quantum operations $\{\Rc_{\Ec,\sigma_t}(\cdot)\}$, i.e.
\begin{equation}Z_{t+1}=g\left(Y_{t+1}\right)\le {\Rc_{\Ec,\sigma_t}}\left(g\left(Y_t\right)\right)={\Rc_{\Ec,\sigma_t}}\left(Z_t\right).
\end{equation}
\end{corollary}
\proofn
Recall that, in view of (\ref{rtspacetimeq}), $Y_t=\sigma_t^{-\um}\rho_t\sigma_t^{-\um}$ is space-time harmonic in reverse time with respect to the quantum operations $\{\Rc_{\Ec,\sigma_t}(\cdot)\}$. Observe that $\sigma_t^{-\um}\rho_t\sigma_t^{-\um}$ has (real) eigenvalues in $(0,+\infty)$.
Observe moreover that the function $g(x)=x\log(x)$ is operator convex on $(0,+\infty)$ (see \cite[Exercise V.2.13]{bhatia}).  The conclusion now follows from a reverse-time version of Proposition \ref{operatorsubmartingale}.
\qed

This can be seen as an H-Theorem in operator form: In fact, as in the classical case, the reverse time subharmonic property of $\{Z_t\}$ of Theorem \ref{qsubm} implies under expectation a more usual, Lindblad-Araki-Uhlmann-like  \cite{lindblad-entropy,araki,uhlmann} form of the $H$-theorem. Namely, we obtain monotonicity for the Belavkin-Staszewski's relative entropy under completely positive, trace-preserving maps. The same result has been derived for conditional expectations in \cite{Hiai-P}. 
\begin{corollary}
Consider two quantum Markov evolutions associated to the initial conditions $\rho_0\neq\sigma_0,$ and  to the same family of TPCP maps $\{\Ec^\dag_t\}$. Assume that  $\rho_t,\sigma_t$ are  invertible for all $t$'s. Then:
\beq\D_{BS}(\rho_{t+1}||\sigma_{t+1})\le \D_{BS}(\rho_t||\sigma_t) .\eeq
\end{corollary}
\proofn Let $Y_t=\sigma_{t}^{-\um}\rho_{t}\sigma_{t}^{-\um}$ as above. By Theorem \ref{qsubm} and (\ref{constantexp}), we get
\begin{eqnarray}\nonumber\D_{BS}(\rho_{t+1}||\sigma_{t+1})&=&\trace\left(\sigma_{t+1}g(Y_{t+1})\right)\\&\leq& \trace\left(\sigma_{t+1}\Rc_{\Ec,\sigma_t}\left(g(Y_t)\right)\right)\\&=&\trace\left(\Rc_{\Ec,\sigma_t}^\dag(\sigma_{t+1})g(Y_t)\right)=\D_{BS}(\rho_{t}||\sigma_{t}).
\end{eqnarray}
\qed
\noindent If $\bar\sigma$ is the unique stationary state of $\{\Ec^\dag_t\},$ we get a quantum version of the second law.

\section*{Acknowledgements}
\noindent
The authors wish to thank Prof. L. Viola for valuable discussions on time-reversal of quantum operations and quantum error correction. Work partially supported by the MIUR-PRIN Italian grant ``Identification and Control of Industrial Systems", by the GNAMPA-INDAM grant ``Teoria del Controllo per Sistemi Quantistici" and by the Department of Information Engineering research project ``QUINTET".


\appendix
\section{ Operator Jensen's Inequalities}\label{jensen}

We recall here some basic definition and results from the theory of majorization for Hermitian operators on $C^*$-algebras, restricted to our finite-dimensional setting. We refer to \cite[Chapter 5]{bhatia} for a thorough discussion on related topics.

Positive, Hermitian matrices are endowed with a natural partial ordering, that is $A\leq B$ if $\langle \phi |A|\phi\rangle\leq \langle \phi |B|\phi\rangle$ for every $|\phi\rangle \in\Hi.$ Since the spectral theorem applies, if the spectrum of $A$ is contained in some interval ${\cal I}\subset \R$, we can define the action of a real function $f:{\cal I}\rightarrow \R$ on Hermitian matrices by standard functional calculus:
$$f(A)=f(\sum_j\lambda_j\Pi_j)=\sum_jf(\lambda_j)\Pi_j,$$
for any $A^\dag=A=\sum_j \lambda_j \Pi_j,\; \lambda_j\in {\cal I}$ for all $j$'s.
A function $f$ is called {\em operator convex} if $f(\lambda A+(1-\lambda)B)\leq \lambda f(A)+(1-\lambda)f(B),$ for any $\lambda\in[0,1],$ and matrices $A,B$ with spectrum in ${\cal I}$. 
Consider now a set of operators $\{M_k\},$ such that $\sum_k M_k^\dag M_k =I.$ Then, for every tuple $\{ X_k \}$ of self-adjoint matrices, the
operator sum $\sum_k M_k^\dag X_k M_k$ can be thought as an ``operator convex combination'' of the $\{ X_k \}.$ Remarkably, an operator analogue of Jensen's inequality \cite{rudin} holds (see \cite{Hansen-P} and reference therein for a review of the literature on the subject). We give here a reduced statement of Theorem 2.1 in \cite{Hansen-P} which is sufficient to our scope.

\begin{theorem}[Operator Jensen's Inequality]\label{operatorjensen} A function $f:{\cal I}\rightarrow \R$ is operator convex if and only if for any Hermitian $X$ and set of operators $\{M_k\}$ such that $\sum_k M_k^\dag M_k =I$ it satisfies
\beq\label{eq:operatorjensen}f(\sum_k M_k^\dag X_k M_k) \leq \sum_k M_k^\dag f(X_k) M_k.\eeq
\end{theorem}

\noindent {Trace Jensen's inequality} then follows as a corollary.

\begin{corollary}[Trace Jensen's Inequality] Let $f:{\cal I}\rightarrow \R$ be operator convex. Then{\em:
\beq\label{tracejensen}\trace \big(f\big(\sum_k M_k^\dag X_k M_k\big)\big) \leq \trace\big(\sum_k M_k^\dag f(X_k) M_k.\big)\eeq }
\end{corollary}

\noindent It can be shown that, for \eqref{tracejensen} to hold, $f$ suffices to be convex \cite{Hansen-P}.  Another version of Jensen's inequality under trace is related to the contractive version of the Jensen's operator inequality (\cite{Hansen-P}, Corollary 2.3), where the requirement on the $\{M_k\}$ is relaxed to $\sum_k M_k^\dag M_k \leq I.$ 

\begin{proposition}[Expectation Jensen's Inequality] Let $f:{\cal I}\rightarrow \R$ be  convex (not necessarily operator convex). Then{\em:
\beq\label{expectationjensen}\trace \big(\rho f(X)\big) \geq f(\trace\big(\rho X\big)).\eeq }
\end{proposition}
%

\section{ Second Law and Maximum Entropy on Quantum Path Space}\label{pathsection}

The theoretical framework and the results we developed in this paper are closely related to maximum entropy problems on path-space we studied in \cite{pavon-bridges}. We recall in the following the main ingredients and the relevant result. 

Consider a quantum Markov evolution for a finite dimensional system $\cal Q$ with associated Hilbert space $\Hi_{\cal Q}$, generated by an initial density matrix $\sigma_0$ and a sequence of TPCP maps $\{\Ec^\dag_t\}_{t\in[0, T-1]},$ with each $\Ec^\dag_t$ admitting a Kraus representation with matrices $\{M_k(t)\}.$ 

We define a set of possible trajectories, or {\em quantum paths}, by considering a time-indexed family of observables $\{X_t\},\,X_t=\sum_{i=1}^{m_t} x_{i}\Pi_{i}(t),$ with $t\in[0,T].$ The paths are then all the possible time-ordered sequences of events  $\left(\Pi_{i_0}(0),\Pi_{i_1}(1),\ldots,\Pi_{i_T}(T)\right),$ with $i_t\in[1,\,m_t].$
The joint probability for a given path is then given by:
\beq \label{weight} w^\Ec_{(i_0,{i_1},\ldots,{i_T})}(\sigma_0) = \trace\left(\Pi_{i_T}(T)\Ec^\dag_{T-1}(\Pi_{i_{T-1}}(T-1)\ldots \Ec^\dag_{0}(\Pi_{i_0}(0)\sigma_0\Pi_{i_0}(0)) \ldots )\Pi_{i_T}(T)\right).\eeq

Consider now a situation in which we have a reference process case where the {\em initial state} is $\sigma_0$ and the transitions are given by $\{\Ec^\dag_t\}.$ We can look for a process with a different initial density matrix which minimizes releative entropy on path space. Assume $X_0$ to have non-degenerate spectra. In \cite{pavon-bridges} we proved the following.

\begin{theorem}\label{qtheo2}A solution to the problem:\\
\begin{equation}\label{QMEs2}{\rm minimize}\quad \D(w^\Fc(\bar\rho_0)\|w^\Ec(\sigma_0)); \end{equation}
with $w^\Fc(\bar\rho_0)$ the path space probability distribution induced by a family of TPCP maps $\{\Fc^\dag_t\}$ and initial state $\bar\rho_0,$
is given by the quantum Markov process with initial density $\bar\rho_0$ and forward transitions: 
\beq\Fc_t(\cdot)=\Ec_t(\cdot),\quad\forall t\in[0,T-1].
\eeq
\end{theorem}

\noindent The total cost then depends only on the initial condition and can be bounded by $\D_{U}(\bar\rho_0\|\sigma_0).$ 
Altough one would expect the problem solution to depend on the choice of the quantum path-space,  {\em it turns out to be independent} from the choice of the observables $\{X_t\}_{t\in[0,T]}.$ 
This shows how, given {\em any path space} and any pair of intial conditions, two quantum Markovian evolutions generate ``trajectories'' that are the closest in relative entropy if they evolve according to the same transition mechanism. In particular, if the reference evolution corresponds to a $\bar{\sigma}$ which is the unique stationary density for the transitions $\{\Ec^\dag_t\},$ Theorem \ref{qtheo2} can also be interpreted as a generalized form of the second law of thermodynamics. Moreover, this result establishes a link between the dissipative behavior of an underlying quantum dynamical system and the classical trajectories associated to any sequence of measurements on the system itself.


\begin{thebibliography}{1}

\bibitem{accardi-qsp} L. Accardi, A. Frigerio and J.T. Lewis. \newblock Quantum stochastic processes, \newblock {\em Publ. Res. Inst. Math. 
Sci.} 18 (1), 97Ð133, 1982. 

\bibitem{araki} H. Araki, Relative entropy for states of von Neumann algebras, {\em Publ. RIMS Kyoto Univ.}, {\bf 11} (1976), 809-833.

\bibitem{barnum-reversing}
H.~Barnum and E.~Knill.
\newblock Reversing quantum dynamics with near-optimal quantum and classical
  fidelity.
\newblock {\em Journal of Mathematical Physics}, 43(5):2097--2106, 2002.

\bibitem{bhatia}
R. Bhatia.
\newblock {\em Matrix Analysis.}
\newblock Springer-Verlag, New York, 1997.

\bibitem{B-S}
V.~P. Belavkin and P.~Staszewski.
\newblock C*-algebraic generalization of relative entropy and entropy.
\newblock {\em Annales de l\'{}Institute Henri Poincar\'{e}}, 37(1):51--58, 1982.

\bibitem{viola-IPS}
R.~Blume-Kohout, H.~K. Ng, D.~Poulin, and L.~Viola.
\newblock The structure of preserved information in quantum processes.
\newblock {\em Physical Review Letters}, 100:030501:1--4, 2008.

\bibitem{bratteli}
O.~Bratteli and D.~Robinson.
\newblock {\em Operator Algebras and Quantum Statistical Mechanics, Volumes I
  and II.}
\newblock Springer-Verlag, Berlin, second edition, 2002.

\bibitem{bremaud} P. Br\'emaud, {\em Markov Chains. Gibbs Fields, Monte Carlo Simulation, and Queues}, Springer-Verlag, New York, 1999.



\bibitem{crooks}
G.~Crooks.
\newblock Quantum operation time reversal.
\newblock {\em Physical Review A}, 77:034101:1--4, 2008.


\bibitem{doob-markov}
J.~L. Doob.
\newblock A markov chain theorem.
\newblock {\em Probability \& Statistics (The H. Cram\'er Volume)}, pages
  50--57, 1959.


\bibitem{F-K}
J.~I. Fuji and E.~Kamei.
\newblock Relative operator entropy in noncommutative information theory.
\newblock {\em Math. Japon.}, 34:341--348, 1989.


\bibitem{Hansen-P}
F.~Hansen and G.~K. Pedersen.
\newblock Jensen's operator inequality.
\newblock {\em Bull. London Math. Soc.}, 35:553--564, 2003.

\bibitem{Hiai-P}
F.~Hiai and D.~Petz.
\newblock The proper formula for relative entropy and its asymptotics in
  quantum probability.
\newblock {\em Communication in Mathematical Physics}, 143:99--114, 1991.

\bibitem{horn-johnson}
R.~A. Horn and C.~R. Johnson.
\newblock {\em {Matrix Analysis}}.
\newblock Cambridge University Press, New York, 1990.

\bibitem{knill-qec}
E. Knill and R. Laflamme.
\newblock Theory of quantum error-correcting codes.
\newblock {\em Physical Review A}, 55(2):900--911, 1997.

\bibitem{kraus}
K.~Kraus.
\newblock {\em States, Effects, and Operations: Fundamental Notions of Quantum
  Theory}.
\newblock Lecture notes in Physics. Springer-Verlag, Berlin, 1983.

\bibitem{kribs-OEC} D.W. Kribs, R.W. Spekkens. \newblock Quantum error correcting subsystems are unitarily recoverable subsystems, \newblock{\em Phys. Rev. A}, 74, 042329, 2006.

\bibitem{kummerer} B. Kummerer. Quantum Markov Processes and Application in Physics,
\newblock in {\em Quantum Independent Increment Processes II},  Lecture Notes in Mathemtics 1866, \newblock Springer Berlin - Heidelberg, 2006.

\bibitem{lindblad-entropy}
G.~Lindblad.
\newblock Completely positive maps and entropy inequalities, 
\newblock {\em Communication in Mathematical Physics}, {\bf 40}, 147--151 1975.

\bibitem{nelson-adjoint}
E.~Nelson.
\newblock The adjoint markov process.
\newblock {\em Duke Math. J.}, 25:671--690, 1958.

\bibitem{N1}E.Nelson,{\em Dynamical
Theories of Brownian Motion}, Princeton
University Press, Princeton, 1967.

\bibitem{neveu} J. Neveu, {\em Discrete-parameter martingales}, North-Holland, Amsterdam; American Elsevier, New York, 1975.

\bibitem{nielsen-chuang}
M.~A. Nielsen and I.~L. Chuang.
\newblock {\em Quantum Computation and Quantum Information}.
\newblock Cambridge University Press, Cambridge, 2002.

\bibitem{NC-rev} M.~A. Nielsen, C. M, Caves, B. Schumacher and H. Barnum.
\newblock Information-theoretic approach to quantum error correction and reversible measurement.
\newblock {\em Proc. R. Soc. Lond. A}, 454, 266-304, 1998.

\bibitem{parthasaraty} K.R. Parthasarathy. {\em An Introduction to Quantum Stochastic Calculus}, Birkh\"{a}user-Verlag, Basel, 1992. 

\bibitem {pavon} M.Pavon, Stochastic control and nonequilibrium thermodynamical systems, {\it Appl. 
Math. and Optimiz.} {\bf 19} (1989), 187-202.

\bibitem{pavon-bridges}
M.~Pavon and F.~Ticozzi.
\newblock Discrete-time classical and quantum  Markovian evolutions: Maximum entropy  problems on path space, April 2009, submitted for publication, preprint arXiv:math-ph/0811.0933v2.

\bibitem{Petz-charofrelent}
D.~Petz.
\newblock Characterization of the relative entropy of states of matrix
  algebras.
\newblock {\em Acta Math. Hung.}, 59:449--455, 1992.

\bibitem{rudin} W. Rudin, {\em Real and Complex Analysis}, McGraw-Hill, 1987.

\bibitem{takesaki}M. Takesaki.\newblock Conditional expectations in von Neumann algebras, \newblock {\em J. Funct. Anal.} {\bf 9} 
306Ð321, 1972. 

\bibitem{uhlmann} A. Uhlmann, Relative entropy and the Wigner-Yanase-Dyson-Lieb concavity in an interpolation theory, {\em Commun. Math. Phys.}, {\bf 54} (1977), 21-32.

\bibitem{Ume}
H.~Umegaki.
\newblock Conditional expectations in an operator algebra iv (entropy and
  information).
\newblock {\em Kodai Math. Sem. Rep.}, 14:59--85, 1962.

\bibitem{vedral-relativeentropy}
V.~Vedral.
\newblock The role of relative entropy in quantum information theory.
\newblock {\em Rev. Mod. Phys.}, 74(1):197--234, 2002.




\end{thebibliography}
\end{document}